\documentclass[11pt,a4paper]{article}
\usepackage{jcappub}
\usepackage{physics}

\usepackage{amssymb,amsfonts}
\usepackage{amssymb} 
\usepackage{amsmath, amsthm} 
\usepackage{amsgen, amsfonts, amsbsy} 
\usepackage{mathrsfs} 
\usepackage{color}
\newcommand{\be}{\begin{equation}}
\newcommand{\ee}{\end{equation}}

\definecolor{pinegreen}{rgb}{0.0, 0.47, 0.44}

\def\theequation{\arabic{section}.\arabic{equation}}

\title{\textbf{New time-dependent solutions of viable Horndeski gravity}}

\author[a]{Reza Saadati,}
\author[b]{Andrea Giusti,}
\author[c]{Valerio Faraoni,}
\author[a]{Fatimah Shojai}

\affiliation[a]{Department of Physics, University of Tehran, 
Kargar North Street, Tehran, Iran 14395/547}

\affiliation[b]{Institute for Theoretical Physics, ETH Zurich,
Wolfgang-Pauli-Strasse 27, 8093, Zurich, Switzerland}

\affiliation[c]{Department of Physics \& Astronomy, Bishop's University, 
2600 College Street, Sherbrooke, Quebec, Canada}

\emailAdd{r.saadati@ut.ac.ir}
\emailAdd{agiusti@phys.ethz.ch}
\emailAdd{vfaraoni@ubishops.ca}
\emailAdd{fshojai@ut.ac.ir}

\abstract{We generate new spherical and time-dependent solutions of viable Horndeski gravity
 by disforming a solution of the Einstein equations with scalar 
field source and positive cosmological constant. They describe dynamical 
objects embedded in asymptotically FLRW spacetimes and contain apparent 
horizons and a finite radius singularity that evolve in time in peculiar 
ways apparently not encountered before in Einstein and ``old'' 
scalar-tensor gravity. }


\arxivnumber{}

\begin{document}
\maketitle
\flushbottom

\section{Introduction}
\label{sec:1}
\setcounter{equation}{0}

There is currently much research on alternative theories of gravity with 
motivation ranging from attempts to explain the current acceleration of 
the universe without an {\em ad hoc} dark energy 
\cite{Capozziello:2003tk, 
Carroll:2003wy} to the realization that, as soon as one tries 
to quantum-correct general relativity (GR), new fields or higher order 
terms in the field equations and new degrees of fredom appear that make 
the resulting theory deviate in essential ways from 
Einstein gravity.

``First generation'' scalar-tensor gravity \cite{Brans:1961sx, 
Bergmann:1968ve, Nordtvedt:1968qs, Wagoner:1970vr, Nordtvedt:1970uv} 
originating with Brans-Dicke theory \cite{Brans:1961sx}, contains only one 
extra scalar degree 
of freedom in addition to the two massless spin two modes of GR and is the 
prototypical alternative gravity. After the 1998 discovery that the 
present expansion of the universe is accelerated, an {\em ad hoc} and very 
exotic ({\em i.e.}, with equation of state parameter $w\simeq
-1$) dark energy was introduced in GR-based cosmology to explain this 
 phenomenon \cite{AmendolaTsujikawabook}. This postulate led to the 
 $\Lambda$-Cold Dark Matter ($\Lambda$CDM) standard model of cosmology. 
 This dark energy introduced overnight to fit observational data is deeply 
 unsatisfactory and has motivated the search for alternative explanations 
 of the cosmic acceleration, reviving the interest in alternative gravity, 
 with the idea that perhaps on large scales gravity is not described by GR 
 but by some alternative theory with extra degrees of freedom 
 \cite{Capozziello:2003tk,Carroll:2003wy, Faraoni:2004pi, 
 Faraoni:2010pgm}. Among several proposals $f(R)$ gravity, which is a 
 subclass of scalar-tensor gravity, has probably become the most popular 
 for this purpose (see Refs.~\cite{Sotiriou:2008rp, 
 DeFelice:2010aj,Nojiri:2010wj} for reviews).

In the last decade, the study of scalar-tensor gravity has gone well 
beyond first generation, Brans-Dicke-like theories, reviving 
\cite{Deffayet:2009wt,Deffayet:2009mn, Deffayet:2011gz}  the old Horndeski 
theory of 
gravity \cite{Horndeski:1974wa}, 
which was believed to be the most general scalar-tensor theory described 
by second order field equations \cite{Deffayet:2009wt, Deffayet:2009mn, 
Deffayet:2011gz}. This belief 
was revised 
when it was discovered that, among higher order scalar-tensor theories 
beyond Horndeski, imposing a special degeneracy conditions makes the field 
equations of second order again \cite{Gleyzes:2014dya, Gleyzes:2014qga, 
Langlois:2015cwa, Langlois:2015skt, 
BenAchour:2016cay, Crisostomi:2016czh, Motohashi:2016ftl, 
BenAchour:2016fzp, Crisostomi:2017aim}. 
These 
theories have come to be 
known as 
Degenerate Higher Order Scalar-Tensor ({\em DHOST}) theories (see 
\cite{Langlois:2018dxi, Langlois:2017mdk} for reviews).

The field equations of Horndeski and {\em DHOST} gravity contain many 
terms and 
are complicated, hence it is difficult to obtain analytical solutions even 
in the presence of symmetries. Therefore, the catalogue of exact solutions 
of the field equations of these theories is rather slim and probably 
comprises geometries that are not of high physical relevance but, until 
our knowledge of analytical solutions expands significantly, one has to 
live with this shortcoming. The difficulty of solving the field equations 
directly has led researchers to use the tool known as disformal 
transformation to generate new solutions from known ones acting as seeds. 
Probably the majority of the known analytical solutions of Horndeski and 
{\em DHOST} gravity that are not already solutions of the coupled 
Einstein-Klein-Gordon equations or of ``old'' scalar-tensor gravity (see 
\cite{Faraoni:2021nhi} for a recent review) have been found using 
disformal transformations, and efforts have gone into assessing the nature 
of the solutions generated by a disformal transformation given the 
properties of the seed solution \cite{BenAchour:2020wiw,Faraoni:2021gdl}, 
following similar discussions for conformal transformations 
\cite{Faraoni:2015paa,Hammad:2018ldj}. The transformation properties of 
the various terms composing the {\em DHOST} action under disformal 
transformations, and their inverses, were analyzed in 
\cite{Achour:2021pla}, while the Petrov classification of geometries 
obtained with disformal solutions of {\em DHOST} theories, and how these 
Petrov 
classes are mapped by disformal transformations, were discussed in 
\cite{BenAchour:2020wiw,Achour:2021pla}. Many of the known solutions are 
stealth \cite{Babichev:2012re, Anabalon:2013oea, Babichev:2013cya,  
Charmousis:2014zaa, Kobayashi:2014eva, Babichev:2016kdt, 
Motohashi:2018wdq} and it is only recently that non-stealth solutions have been 
found \cite{Babichev:2017guv, Babichev:2013cya, Anson:2020trg, BenAchour:2020fgy, 
Chatzifotis:2021hpg}.

Given a solution $\left( g_{ab}, \phi \right) $ of GR with scalar 
field as the matter source, or of scalar-tensor gravity, a disformal transformation maps the 
metric $g_{ab} $ into a new one according to \cite{Bekenstein:1992pj, 
Ezquiaga:2017ner, Zumalacarregui:2010wj, Zumalacarregui:2013pma} 
\be
g_{ab} \rightarrow \bar{g}_{ab} = \Omega^2 \left(\phi , X \right) g_{ab} 
+ W\left( \phi, X \right) \nabla_a \phi \nabla_b \phi  \label{disftransf}
\ee
where $\Omega $ and $W$ are, in principle, functions of $\phi $ and of $
X \equiv  -\frac{1}{2} \, \nabla^c \phi \nabla_c \phi  $.
The conditions 
\be
\Omega \neq 0 \,, \quad \quad \Omega^2 -X\left( \Omega^2\right)_{, X} -X^2 
W_{, X} \neq 0   \label{conditions}
\ee
must hold to ensure invertibility of the map~(\ref{disftransf}) 
\cite{BenAchour:2020wiw}.

A disformal transformation maps solutions of the coupled 
Einstein-Klein-Gordon equations into {\em DHOST} solutions 
\cite{BenAchour:2020wiw,Achour:2021pla}. Here instead we generate a new family of 
solutions of viable Horndeski gravity by disforming 
a solution of the Einstein equations with positive cosmological constant 
sourced by a minimally coupled scalar field. Apart from cosmology, most of 
the known geometries  describing 
spherically  
or axially symmetric objects in {\em DHOST} and Horndeski gravity are 
static or stationary, therefore we 
attempt to go beyond this restriction to learn more about the nature of 
gravity at least in the sub-class of Horndeski gravity compatible with a luminal propagation 
of gravitational waves. Already in GR and in old scalar-tensor theory, 
moving from stationary to time-dependent objects reveals new and richer 
phenomenology. For example, black hole event horizons 
cease being relevant and are replaced by time-dependent apparent horizons 
which, unfortunately, depend on the foliation \cite{Wald:1991zz, 
Schnetter:2005ea}.\footnote{The situation is not always so dire, though: 
for example, in spherical symmetry all spherical foliations give the same 
apparent horizons \cite{Faraoni:2016xgy}.} Time-dependent apparent 
horizons usually appear and annihilate in pairs 
\cite{Husain:1994uj,Booth:2005qc,Nielsen:2005af, Faraoni:2015ula} 
and singularities can be dynamical. 
Most studies of non-asymptotically flat and dynamical analytical 
geometries and of their properties seem to have concentrated on 
spherical objects embedded in Friedmann-Lema\^itre-Robertson-Walker 
(FLRW) universes \cite{Husain:1994uj,Fonarev:1994xq,Faraoni:2015ula}.

In this work, we adopt as seed for the disformal transformation a 
spherically symmetric, time-dependent, and asymptotically FLRW solution of 
the Einstein equations with positive cosmological constant and a scalar 
field with exponential potential as the matter source. It is a special 
case of the Fonarev solution \cite{Fonarev:1994xq} for exponential scalar 
field potential, with the peculiarity that the geometry is dynamical while 
the scalar field $\phi$ is static, which makes the disformal 
transformation to Horndeski gravity rather manageable (a similar 
situation occurs for stealth solutions \cite{Faraoni:2021nhi}). Moreover, 
the geometry is asymptotically FLRW, which distinguishes it from the 
asymptotically flat solutions populating the (still slim) catalogue of 
analytical solutions of Horndeski and {\em DHOST} gravity 
\cite{Faraoni:2021nhi}.

Two solutions are generated in the next section, using two different 
disformal transformations, and their apparent horizons and singularities 
are discussed. Section~\ref{sec:3} analyses them in more detail;  
Sec.~\ref{sec:4} investigates the dynamics of the apparent horizons, while 
Sec.~\ref{sec:5} contains the conclusions.

\textbf{Notation:} We follow the conventions of Ref.~\cite{Waldbook}: the 
signature of the metric tensor $g_{ab}$ is $({-}{+}{+}{+})$, $G$ is 
Newton's constant, and units are used in which the speed of light $c$ is 
unity. Furthermore, again following \cite{Waldbook}, we employ the 
``abstract index notation''. Hence, quantities involving Latin letters will represent
tensorial objects, whereas the presence of Greek letters will denote the choice of a specific
chart.

\section{Disforming a special case of the Fonarev solution}
\label{sec:2}
\setcounter{equation}{0}

The Einstein-scalar field equations admit the 3-parameter family of 
Fonarev solutions  given by \cite{Fonarev:1994xq,Maeda:2007bu}
\be
ds_{\rm F} ^2 = - \mbox{e}^{8\alpha^2 at} A^{\delta}(r) \, dt^2 +\mbox{e}^{2at} 
\left[ 
\frac{dr^2}{A^{\delta} (r) } + A^{1-\delta} (r) r^2 d\Omega_{(2)}^2 
\right] \,,  
\label{Fonarev metr}
\ee
\be
\phi(t,r) = \frac{1}{2\sqrt{\pi}} \left[ 2\alpha \, a t + 
\frac{1}{2\sqrt{1+4\alpha^2}} \, \ln A(r) \right] \,, \label{phi}
\ee
where $d\Omega_{(2)}^2 \equiv d\vartheta^2 +\sin^2\vartheta \, d\varphi^2$ 
is the line element on the unit 2-sphere, 
\begin{eqnarray}
A(r) &=& 1-\frac{2m}{r} \,,\\
&&\nonumber\\
\delta &=& \frac{2\alpha}{\sqrt{1+4\alpha^2}} <1 \,,
\end{eqnarray}
$r\geq 2m$, and where $m, \alpha$, and $a$ are constants ($m>0$ has 
the dimensions of mass, $a$ those of an inverse length, and $\alpha$ is 
dimensionless), while the scalar field is subject to the potential 
\cite{Fonarev:1994xq} 
\be
V(\phi)=V_0 \, \mbox{e}^{-8\sqrt{\pi} \, \alpha \phi } \,, \quad \quad 
V_0 = \frac{ a^2 \left( 3-4\alpha^2 \right)}{8\pi} \,.
\ee
The condition $V \geq 0$ places the restriction $|\alpha| \leq \sqrt{3}/2$ 
on the dimensionless parameter $\alpha$.

The Fonarev solution is spherically symmetric, time-dependent, and 
asymptotically FLRW. It describes a time-dependent wormhole or a naked 
singularity embedded in a FLRW universe (see  
Refs.~\cite{Fonarev:1994xq,Maeda:2007bu,Faraoni:2015ula} for discussions). For 
$\alpha=\pm \sqrt{3}/2$, 
the potential $V(\phi)$ vanishes and the Fonarev spacetime 
reduces to the Husain-Martinez-Nu\~nez solution of the Einstein 
equations with a free scalar field \cite{Husain:1994uj}. In the limit 
$a=0$, it reduces instead to the Fisher-Janis-Newman-Winicour-Wyman static 
and spherical scalar field geometry \cite{Fisher:1948yn, 
Janis:1968zz, Wyman:1981bd, Faraoni:2015ula, Faraoni:2021nhi}.  Here we 
use as a seed 
solution the special case $\alpha=0$ of the Fonarev spacetime, which has a 
time-dependent and asymptotically FLRW geometry but static scalar field:
\be
d\bar{s}^2 = - dt^2 +\mbox{e}^{2at} \left[ dr^2 + A(r) r^2 d\Omega_{(2)}^2  
\right] \,,\label{metricspecialFonarev}
\ee
\be
\phi(r) = \frac{1}{4\sqrt{\pi} }  \, \ln A(r) \,. 
\label{phispecialFonarev}
\ee
In this limit, $\delta=0$ and the scalar field potential degenerates into 
a constant. Therefore, Eqs.~(\ref{metricspecialFonarev}) and 
(\ref{phispecialFonarev}) describe a 2-parameter family of solutions of 
the Einstein equations 
with positive cosmological constant $\Lambda=8\pi V_0 = 
3a^2$ and sourced by a free scalar field, previously reported in 
Eqs.~(2.27) and 
(2.28) of Ref.~\cite{Faraoni:2017afs}.  The areal radius is $\bar{R}(t,r)= 
\mbox{e}^{at} \, r \sqrt{A(r)} $ and $r \geq 2m$ corresponds to real 
values 
of the physical radius $\bar{R}$, with $r=2m$ equivalent to $\bar{R}=0$. 
The scalar field diverges as 
$\bar{R}\rightarrow 0^+$ (or $r \to 2m^{+}$), which is therefore a physical 
singularity of the theory, except when a wormhole throat is present at 
a positive radius $\bar{R}$ \cite{Fonarev:1994xq,Maeda:2007bu}.

Since $\phi$ depends only on the radial coordinate $r$ while $g_{ab}$ 
depends also on time, this Einstein-scalar field solution is particularly 
well-suited to generate a new family of solutions of viable Horndeski gravity 
by means of the disformal transformation
\be
\bar{g}_{ab} \rightarrow {g}_{ab} = \Omega^2 \left(\phi , \bar{X} \right) \bar{g}_{ab} + 
W\left( \phi, \bar{X} \right) \bar{\nabla}_a \phi \bar{\nabla}_b \phi  
\ee
with $\bar{g}_{ab}$ denoting the metric tensor associated with 
\eqref{metricspecialFonarev}, $\bar{\nabla}$ denotes the connection 
compatible with the metric $\bar{g}_{ab}$, while 
\be
 \bar{\nabla}_{\mu} \phi =  
\frac{m}{2\sqrt{\pi} \, Ar^2} \, \delta_{\,\, \mu}^r
\ee
and 
\be
\bar{X} \left( t, r \right) \equiv - \frac{1}{2} \, \bar{g}^{ab} \, \bar{\nabla}_a \phi \bar{\nabla}_b \phi 
= - \frac{m^2 \, \mbox{e}^{-2at}}{ 8\pi r^4 A^2(r)} 
\,.
\ee

The new line element obtained from the disformal transformation reads
\begin{eqnarray}
d{s}^2 &=& -\Omega^2 dt^2 +\left( \mbox{e}^{2at} \Omega^2 
+\frac{Wm^2}{4\pi  A^2 r^4} \right) dr^2 + \Omega^2 \, 
\mbox{e}^{2at} A  r^2 d\Omega_{(2)}^2 \,.
\label{eq:disformedmetricmannaggia}
\end{eqnarray}
If the functions $\Omega$ and $W$ depend only on $\phi$, then 
$\Omega=\Omega(r)$ and $W=W(r)$. In the following we choose
\be
\Omega(\phi) = \frac{1}{ \sqrt{A(r)} } = \mbox{e}^{-2\sqrt{\pi} \, \phi } 
\,,
\ee
then the areal radius of the metric ${g}_{ab}$ associated with the line 
element \eqref{eq:disformedmetricmannaggia} reads 
\be
R(t,r) = r \, \mbox{e}^{at} \,.
\ee
Furthermore, set 
\be
W(r) \equiv  \frac{4\pi r^2 A(r)}{m^2} \, f(r) \,,
\ee
where the (yet to be specified) function $f(r)$ has the dimensions 
of  a length squared. Inverting the 
expression of $\phi(r)$ yields 
\be
 r=\frac{2m}{ 1-\mbox{e}^{4\sqrt{\pi} \, 
\phi} }
\ee
and 
\be
W( \phi) = \frac{16\pi \, \mbox{e}^{4\sqrt{\pi} \, \phi} }{ \left( 1- 
\mbox{e}^{4\sqrt{\pi} \, 
\phi } \right)^2} \, f\left( \frac{2m}{1-\mbox{e}^{4\sqrt{\pi} \, \phi} } 
\right)\,;
\ee 
this choice of $\Omega$ and $W$ automatically satisfies the 
conditions~(\ref{conditions}). 

The disformed line element then reads
\be
d{s}^2 =-\frac{dt^2}{A(r)} + \left[ \mbox{e}^{2at} +\frac{f(r)}{r^2} 
\right] 
\frac{dr^2}{A(r)} + R^2 d\Omega_{(2)}^2 \label{choice}
\ee
with $r>2m$ and $ \mbox{e}^{2at} +f(r)/r^2 >0$.  
If $f(r)/r^2 \rightarrow 0$ as $r\rightarrow +\infty$, the geometry is 
asymptotically de Sitter with comoving time $t$ and Hubble constant 
$H=a=\sqrt{\Lambda/3}$ for $r\rightarrow +\infty$. The scalar field 
$\phi$, which is left unchanged by the disformal transformation, still 
diverges as $r\rightarrow (2m)^{+}$ (or $R\rightarrow 2m \, 
\mbox{e}^{at}$).

\subsection{Curvature coordinates}

We can rewrite the line element~(\ref{choice}) in curvature coordinates 
employing the areal radius $R(t,r) = r \, \mbox{e}^{at} $. Substitution of 
the  relation between differentials 
\be
dr= \mbox{e}^{-at} \left( dR-aRdt\right)
\ee
into Eq.~(\ref{choice}) gives
\begin{eqnarray}
d{s}^2 &=& -\left[ 1- a^2R^2 \left( 1+\frac{ \mbox{e}^{-2at} 
f(r)}{r^2} 
\right) \right] \frac{dt^2}{A(r)} -\frac{2aR}{A(r)} \left( 1+ \frac{ 
\mbox{e}^{-2at} f(r)}{r^2} \right) dt dR \nonumber\\
&&\nonumber\\
&\, & + \left( 1+ \frac{ \mbox{e}^{-2at} f(r)}{r^2} \right) \, 
\frac{dR^2}{A(r)} +R^2 d\Omega_{(2)}^2 \,. \label{mixed}
\end{eqnarray}   
The cross-term in $dtdR$ can be eliminated, and the line element 
diagonalized, by introducing a new time coordinate $T$ defined by
\be
dT=\frac{1}{F} \left( dt +\beta dR\right) \,, \label{dT}
\ee 
where $\beta (t, R) $ is a function to be determined and $F(t, R)$ is an 
integrating factor guaranteeing that $dT$ is an exact differential and 
satisfying
\be
\frac{ \partial}{\partial R} \left( \frac{1}{F} \right) = 
\frac{\partial}{\partial t} \left( \frac{\beta}{F} \right) \,.\label{F}
\ee

Substituting $ dt=FdT -\beta dR$ in the line element~(\ref{mixed}), one 
obtains
\begin{eqnarray}
d {s}^2 &=& - \frac{F^2}{A(r)} \left[ 1-a^2R^2 \left( 1+ \frac{ 
\mbox{e}^{-2at} 
f}{r^2} \right) \right] dT^2\nonumber\\
&&\nonumber\\
&\, & +\frac{2F}{A(r)} \left\{ \beta\left[ 1-a^2R^2 \left( 1+\frac{ 
\mbox{e}^{-2at} f}{r^2} \right) \right] 
-aR\left( 1+ \frac{\mbox{e}^{-2at} f}{r^2} \right) \right\} dTdR 
\nonumber\\
&&\nonumber\\
&\, & 
+ \left\{ -\beta^2 \left[ 1-a^2R^2 \left( 1+ \frac{ \mbox{e}^{-2at} 
f}{r^2}\right) \right] +2aR \beta  \left( 1+ \frac{ \mbox{e}^{-2at} 
f}{r^2} \right) +  \left( 1+ \frac{ \mbox{e}^{-2at} f}{r^2} \right) 
\right\}  \frac{dR^2}{A(r)} \nonumber\\
&&\nonumber\\
&\, & +R^2 d\Omega_{(2)}^2 \,.  
\end{eqnarray} 
By choosing
\be
\beta (t,R) = \frac{ aR\left( 1+ \mbox{e}^{-2at} f/r^2\right) }{1-a^2R^2 
\left( 
1+ \mbox{e}^{-2at} f/r^2\right)}  \label{beta}
\ee
the line element is diagonalized, becoming
\begin{eqnarray}
d{s}^2 &=& -  \left[ 1-a^2R^2 \left( 1+ \frac{ \mbox{e}^{-2at} 
f}{r^2} \right) \right] \frac{F^2 dT^2}{A(r)} \nonumber\\
&&\nonumber\\
&\, & + \frac{ \left( 1+  \mbox{e}^{-2at}f/r^2 \right) }{
A(r) \left[ 1 - a^2R^2 \left( 1+ \, \mbox{e}^{-2at}f/r^2 \right)\right] }
 \, dR^2  +R^2 d\Omega_{(2)}^2 \,. \label{acci}
\end{eqnarray}
 Although $f(r)/r^2$ and $A(r) =1-2m/r=1-2m \, \mbox{e}^{at} /R $ can be 
expressed in terms of $R$ using $r= R\, \mbox{e}^{-at}$ to eliminate $r$, 
the form~(\ref{acci}) of the line element remains implicit because it 
contains the old time coordinate $t$. From the physical point of 
view, however, it is more interesting to descrive the evolution in terms 
of the comoving observers of the asymptotic FLRW background, who use $t$ 
as their time coordinate.

\subsection{Apparent horizons}

The apparent horizons of a spherically symmetric 
spacetime are located by the roots of the equation 
\be
\nabla^c R \nabla_cR=0 \,,\label{AHeq}
\ee
where $R$ is the areal radius ({\em e.g.}, 
\cite{Abreu:2010ru, Faraoni:2015ula}).  
Single roots 
correspond to black hole (or 
possibly, white hole or cosmological) apparent horizons, while double 
roots correspond to wormhole throat horizons 
\cite{Abreu:2010ru, Faraoni:2015ula}.   

For the line element~(\ref{choice}), this equation becomes 
\begin{eqnarray}
\nabla^cR \nabla_cR &=& 
{g}^{tt} \, \left( \frac{\partial R}{\partial t} \right)^2  + 
{g}^{rr} \left( \frac{\partial R}{\partial r} \right)^2 \nonumber\\
&&\nonumber\\
&=& \mbox{e}^{2at} r^2 A(r) \left[ \frac{1}{r^2 \, 
\mbox{e}^{2at} 
+f(r)} -a^2\right]=0 \,.  \label{eq:2.26}
\end{eqnarray}
The first root, which is a single root and always exists, is given by 
$r_1=2m$ ({\em i.e.}, $A(r) = 0$) or 
\be
R_1(t)=2m  \, \mbox{e}^{at} \,.
\ee 
If $a>0$, then $R_1(t) \rightarrow +\infty$ as 
$t\rightarrow +\infty$. The other roots, if they exist, are the real and 
positive solutions of 
\be
r^2 \, \mbox{e}^{2at} +f(r)-\frac{1}{a^2}=0 \label{minchia}
\ee
and they depend on the choice of the function $f(r)$ in the disformal 
transformation. 

\subsection{Singularity}
\label{sec:singul}

Computing the invariants of the Ricci tensor for the line element 
\eqref{choice} for a generic function $f(r)$ yields
\begin{eqnarray}
{\cal R} &=& \frac{4 A(r) \left[  r f'(r)-f(r) \left( a^2 
f(r)+2\right) \right] + r A'(r) \left[ 2 f(r)-r f'(r)\right]}{2 \left[ 
r^2 \, \mbox{e}^{2 a t}+f(r)\right]^2} \nonumber\\ 
&&\nonumber\\
&\, & +\frac{-2 A^2(r) \left[ 2 a^2 f(r) + 1 \right] + r^2 
A(r) A''(r) - r^2 \left[ A'(r) \right]^2}{A(r) \left[ r^2 
\, \mbox{e}^{2 at} + f(r)\right]} + 12 a^2 A(r)\nonumber\\
&&\nonumber\\
&\, & + \frac{2 \, \mbox{e}^{-2 a t}}{r^2} \,, \label{ricci}\\
&&\nonumber\\
{\cal R}_{ab} {\cal R}^{ab} &=& \frac{1}{16 \left( r^2 \, \mbox{e}^{2 at}
+ f\right)^4} \,  
\Bigg\{ -32 a^2 \left[ r^2 \, \mbox{e}^{2 at} + f\right]  
\Bigg[ r  A'(r) \left[ r^2 \, \mbox{e}^{2 a t} + f\right] 
+ 2 A(r) f \Bigg]^2  \nonumber\\
&&\nonumber\\
&\, & +\frac{1}{r^4}\Bigg[ 8 \, \mbox{e}^{-4at} 
\Big\{ r^3 \, \mbox{e}^{2at}  \left[ A(r) \left(6 a^2 r^3 \, \mbox{e}^{4at}
-2 r \, \mbox{e}^{2at} + f' \right) +2r  \, \mbox{e}^{2 a t}\right] 
\nonumber\\
&&\nonumber\\
&\, & + 2 r^2 \, \mbox{e}^{2at} f \left[ A(r) \left( 5 a^2 r^2 \, 
\mbox{e}^{2at} - 2\right) + 2\right]  \nonumber\\
&&\nonumber\\
&\, & +f^2 \left(4 a^2 r^2 \, \mbox{e}^{2 a t} A(r) +2 \right) 
\Big\}^2 \Bigg] \nonumber\\
&&\nonumber\\
&\, & + \frac{1}{A^2(r)} \Bigg[ \Big(4 a^2 A^2(r) \left(2 f \left(3 
r^2 \, \mbox{e}^{2 at} + f \right) + 3 r^4 \, \mbox{e}^{4 a t}\right)
-2 r^2 (A'(r))^2 \left[ r^2 \, \mbox{e}^{2 at} + f\right] \nonumber\\
&&\nonumber\\
&\, & +r A(r) \left(2 r A''(r) \left(r^2 \, \mbox{e}^{2 a 
t}+f\right)+A'(r)  
\left(4 r^2 \, \mbox{e}^{2at} -r f' + 6\right)\right) \Big)^2 \nonumber\\
&&\nonumber\\
&\, &  +\Big[ 4 A^2(r) 
\left(f \left(4 a^2 r^2 \, \mbox{e}^{2at}-2\right)+3 a^2 r^4 \, 
\mbox{e}^{4at} +r f'\right] - 2 r^2 A'(r)^2 \left(r^2 \, \mbox{e}^{2 a t} 
+ f\right)  \nonumber\\
&&\nonumber\\
&\, & + r A(r) 
\left(2 r A''(r) \left(r^2 \, \mbox{e}^{2at} + f \right)-A'(r) \left(r 
\left(4 r \, \mbox{e}^{2at} + f'  \right)+2 
f\right)\right)\Big)^2\Bigg] \Bigg\} \,.  \label{ricci2}
\end{eqnarray}

\begin{figure}[h] 
{\includegraphics[width=0.48\textwidth]{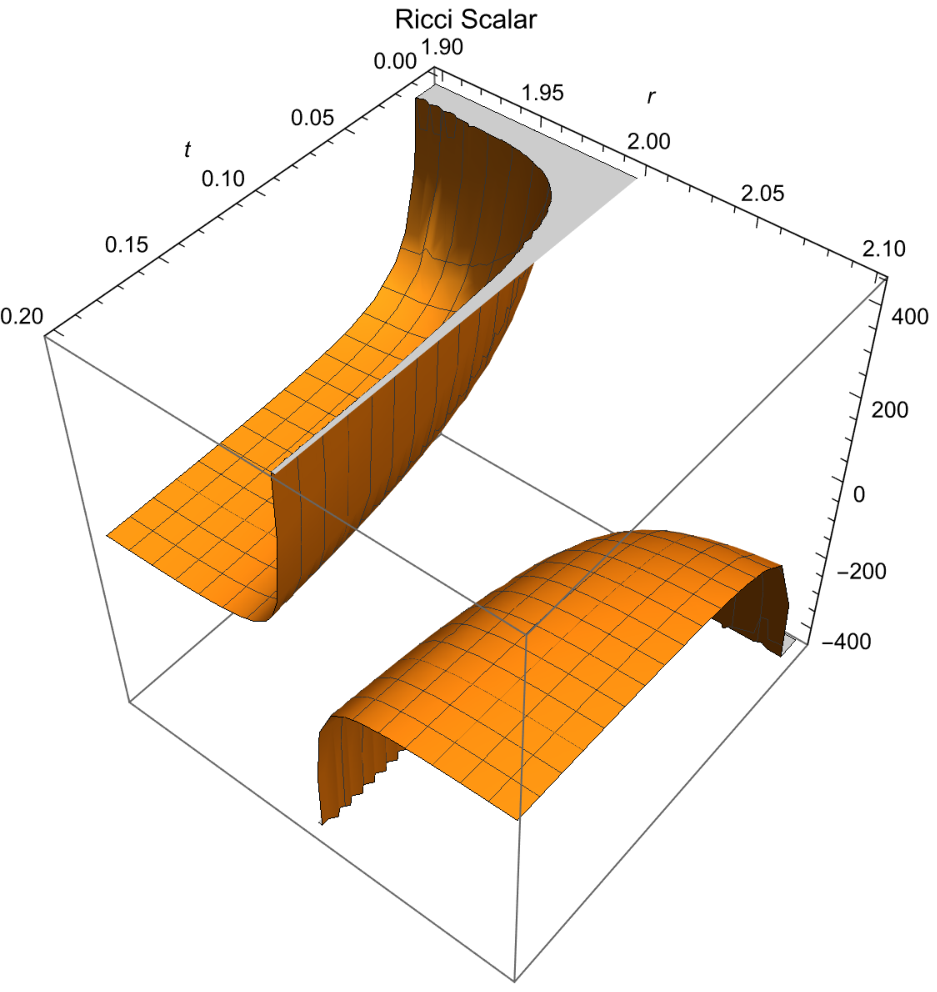}}
{\includegraphics[width=0.48\textwidth]{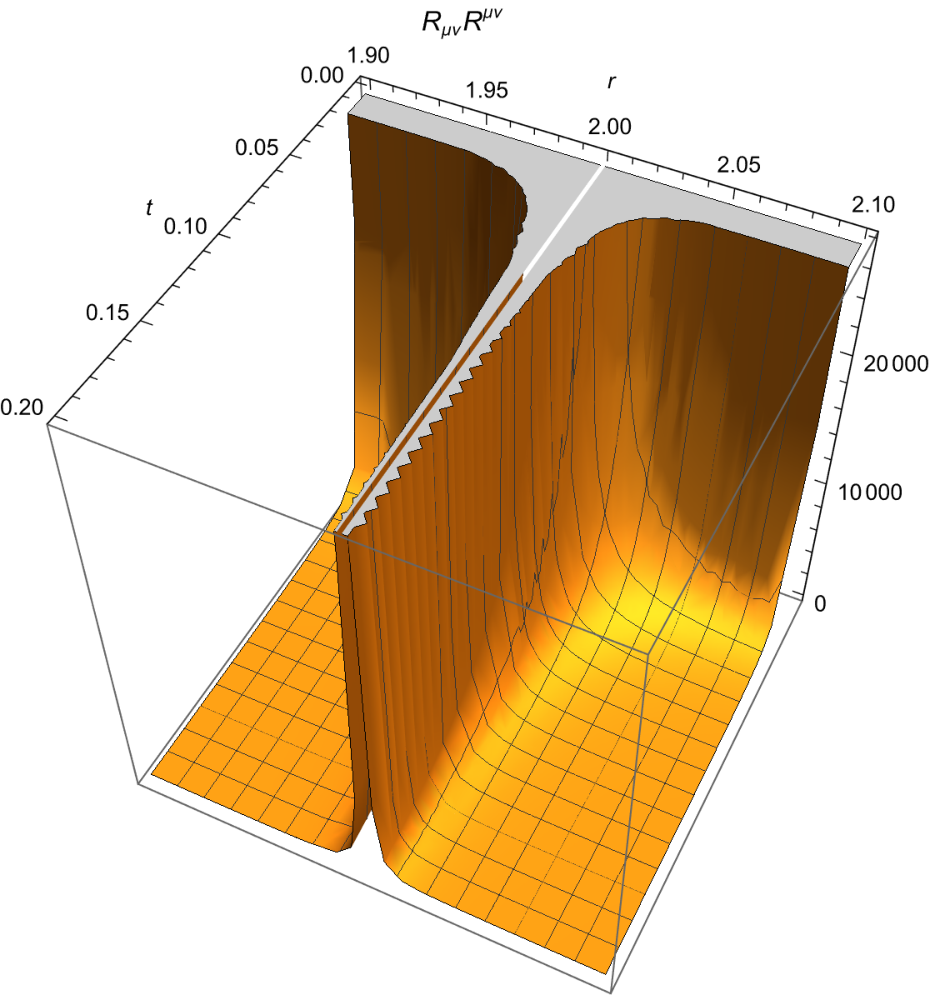}}
\caption{\label{fig:riccis} The Ricci scalar~(a) and Ricci 
tensor squared~(b) 
for the parameter values $a=m=1$. These invariants diverge at $r=2m$ for 
all values of $t$.}
\end{figure}

One can immediately conclude that $r_1=2m$ ({\em i.e.}, $A(r) = 0$) is a curvature singularity for \eqref{choice} since both ${\cal R}$ and ${\cal R}_{ab} {\cal R}^{ab}$ 
diverge there, together with the scalar field~(\ref{phi}). Note however that as $r\to (2m)^+$ one finds
\begin{equation}
{\cal R} \simeq  -\frac{1}{A(r) [r^2 \, \mbox{e}^{2 a t}+f(r)]} 
\,,\label{qq1}
\end{equation}
\begin{equation}
{\cal R}_{ab} {\cal R}^{ab} \simeq 
\frac{1}{A^2(r)[r^2 \, \mbox{e}^{2 a t}+f(r)]^2}  \,. \label{qq2}
\end{equation}
Unless the denominators in the right-hand sides of Eqs.~(\ref{qq1}) 
and~(\ref{qq2}) diverge (and they diverge at different rates), 
these Ricci invariants are certainly divergent. For instance, 
for $f(r)=-r^2$ the geometry has a curvature singularity at 
$r_1=2m$. Figure~\ref{fig:riccis} illustrates  these divergencies for the 
parameter values  $a=m=1$.

Furthermore, other curvature singularity can emerge from the roots of $ 
\mbox{e}^{2at} +f(r)/r^2 =0$, if $r\neq 2m$.

The singularity surface $r=2m$ of the scalar $\phi$ in 
\eqref{phispecialFonarev}, and consequently of the line element 
\eqref{choice}, is null. In fact, it is defined by the equation $\psi(r) 
\equiv r-2m =0 $ and the normal vector $N^a$ has components
\be
N_\mu \equiv \nabla_\mu \psi =\delta^r _{\,\,\, \mu}
\ee 
and norm squared 
\be
N_\mu N^\mu= \nabla^\mu\psi\nabla_\mu\psi = g^{rr}= \frac{A(r)}{ \mbox{e}^{2at} 
+f(r)/r^2} \,,
\ee
which vanishes as $r \to 2m^{+}$, therefore $r=2m$ is a null surface.

\section{Two simple choices of $f(r) $}
\label{sec:3}
\setcounter{equation}{0}

Here we consider two concrete choices of the free function $f(r)$ that 
produce relatively simple, yet interesting, geometries.

\subsection{Choice $f(r)=0$}

If $f(r)=0$, then $W=0$, and the transformation~(\ref{disftransf}) 
reduces to a pure conformal transformation. This case describes a finite 
radius singularity embedded in a de Sitter universe, which is 
locally static below its de Sitter horizon \cite{Schleich:2009uj}.

The transformed line element is
\be
d{s}^2= -\frac{dt^2}{A(r)} + \frac{ \mbox{e}^{2at} }{A(r)} \, dr^2  
+R^2 d\Omega_{(2)}^2 \,; 
\ee
to express it in terms of the areal radius, note 
that Eq.~(\ref{beta})  gives 
\be
\beta(R) = \frac{aR}{1-a^2R^2} \label{beta2}
\ee
and then Eq.~(\ref{F}) for the integrating factor $F(t,R) $ reduces to 
$\partial_R F = \beta \, \partial_t F$, which admits the constant solution 
$F=1$ and turns the line element into 
\begin{equation}
d{s}^2 = - \frac{\left( 1-a^2R^2\right)}{A(r)} \, dT^2 + 
\frac{1}{\left( 1-a^2 R^2 \right)A(r) } dR^2+R^2 d\Omega_{(2)}^2 
\end{equation}
for $ 2m \, \mbox{e}^{at}< R<1/a$, where the lower bound on the physical 
radius is the location of the curvature singularity. This relation 
implies that it must be $am 
< 1/2$. The transformation between old and new 
time coordinates can be found explicitly in this case: using $F=1$ 
and~(\ref{beta2}), Eq.~(\ref{dT}) is integrated to
\be
T(t,R) = t -\frac{1}{2a} \ln \left( 1-a^2R^2\right) + \mbox{const.}
\ee

The formal (single) root $R=1/a$ of the equation $\nabla^c R \nabla_c 
R=g^{RR}=0$ is reminiscent of the de Sitter horizon of the de Sitter space 
in which the central object is embedded. (If $m=0$, this line element 
reduces to the de Sitter one in static coordinates, with Hubble constant 
$H=a$.) Accordingly, this apparent horizon is always a null surface. In 
fact, the normal vector to the 
apparent horizon defined by $\psi = R - a^{-1} = 0$ reads 
\begin{equation}
N_\mu \equiv \nabla_{\mu}\psi 
= \delta^{R}_{\,\,\,\mu} 
\end{equation}
has norm squared 
\be
N^\alpha N_\alpha = g^{RR} =0 \quad \mbox{on} \,\, R = a^{-1} \,.
\ee 
Since $ 2m \, \mbox{e}^{at}< R<1/a$, the physical spacetime region 
shrinks as the lower bound $ 2m \, \mbox{e}^{at}$ to the radius $R$ grows 
exponentially while the upper bound $1/a$ remains constant.

\subsection{Choice $f(r)= -r^2$}

In the following we assume that $a>0$. If $f(r)=-r^2$, 
\be \label{metr}
W(\phi) = -\frac{ 64\pi m^2 \, \mbox{e}^{4\sqrt{\pi} \, \phi} }{\left( 
1-\mbox{e}^{4\sqrt{\pi} \, \phi } \right)^4} 
\ee
and the disformed line element 
\begin{eqnarray}
d{s}^2 &=& -\frac{dt^2}{A(r)} +\left( \mbox{e}^{2at} -1 \right) 
\frac{dr^2}{A(r)} 
+R^2 d\Omega_{(2)}^2 \\
&&\nonumber\\
&=& 
-\frac{ \left[ 1-a^2R^2 \left( 1-\mbox{e}^{-2at} \right)\right]}{A(r)} \,  
F^2  dT^2  + \frac{ \left( 1-\mbox{e}^{-2at}\right)dR^2}{A(r) \left[ 
1-a^2R^2 \left( 1-\mbox{e}^{-2at} \right) \right]} +R^2 d\Omega_{(2)}^2 
\end{eqnarray}
is defined for $t\geq 0$ and $r>2m$, equivalent to $R> R_1(t) =2m 
\, \mbox{e}^{at} $. The geometry is not asymptotic 
to a FLRW universe for large $r$, however at late times {\em and} large 
$r$ it approaches a de Sitter space with scale factor $ \mbox{e}^{at}$. 

The singularity is, again, described by the equation $\psi \equiv r-2m=0$.
As discussed in Sec.~\ref{sec:singul}, this is a null singularity somehow 
similar to the thunderbolt singularities discussed in the 
literature~\cite{Hawking:1992ti, Ishibashi:2002ac,Misonoh:2015nwa}.

\begin{figure}[h]
\includegraphics[width=0.6\textwidth]{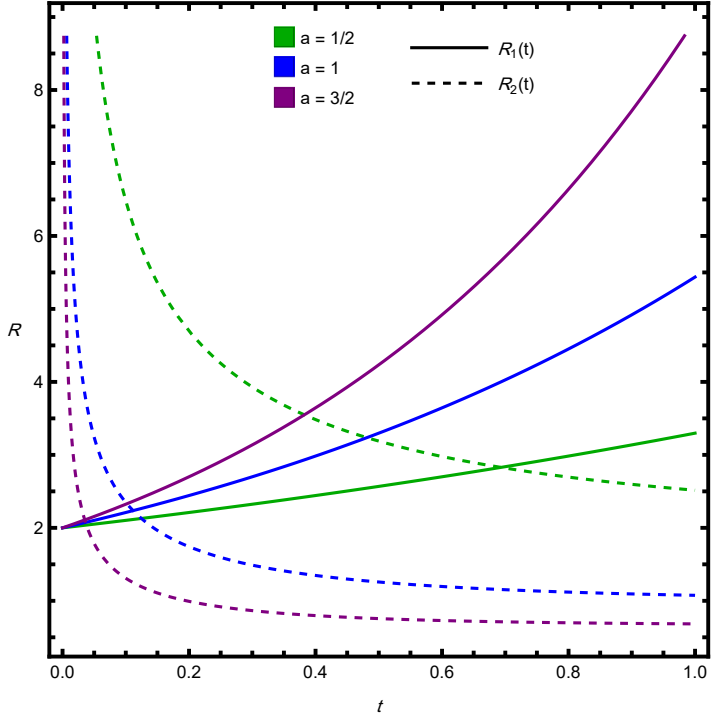}
\caption{\label{fig:f(r)=-r^2} The behaviour of the physical radii of the 
singularity (solid curves) and of the apparent horizon (dashed curves) 
for  $f(r)=-r^2$ and for three values of the parameter $a$.}
\end{figure}

Let us describe the dynamics of the singularity and apparent 
horizon in terms of the time coordinate $t$, which is the physical time 
of the 
comoving observers of the background space (that reduces asymptotically to 
a de Sitter universe in the limits discussed). Equation~(\ref{eq:2.26}) 
admits the single root $R_1(t)=2m \, \mbox{e}^{at}$, which 
describes an 
expanding singularity, plus the other single root
\be \label{horizon1}
r_2(t) = \frac{\mbox{e}^{-at/2} }{a \sqrt{ 2\sinh (at)}} 
= \frac{1}{a \, \sqrt{ \mbox{e}^{2at}-1}} 
\,. 
\ee
The physical radius of this apparent horizon is
\be
R_2(t) = \frac{ \mbox{e}^{at/2} }{a \sqrt{ 2\sinh (at)}} = 
\frac{ \mbox{e}^{at} }{ a \sqrt{ \mbox{e}^{2at} -1 } } \,;
\ee
it converges to the de Sitter horizon $ H^{-1}= 1/a $ of the de 
Sitter ``background'' from above as $t\rightarrow +\infty$. As 
$t\rightarrow 0^{+}$, $R_1 \rightarrow 2m$ and $R_2\rightarrow +\infty$. 
As time progresses from $t=0$, $R_1(t)$ increases exponentially  
while $R_2(t) $ decreases monotonically from infinity since 
\be
\dot{R}_2= -\frac{ \mbox{e}^{-2at}}{\left( \, \mbox{e}^{2at} -1 
\right)^{3/2} } <0 
\ee
and formally 
approaches the value $1/a$ as $t\rightarrow +\infty$. The 
expanding singularity and the shrinking apparent 
horizon meet at the critical time
\be
t_* =  \frac{1}{2a} \, \ln \left( \frac{4a^2m^2 + 1}{4a^2m^2} \right) >0 
\,,
\ee
after which $R_1$ becomes larger than $R_2$. Since it must be 
$R>R_1$ at all times, this means that the apparent horizon of  
radius $ R_2(t) $ disappears from spacetime at the time $t_*$.  The 
physical picture emerging from these considerations is the following 
(Fig.~\ref{fig:f(r)=-r^2}):

\begin{itemize}

\item For $0< t <t_*$ there is a finite radius singularity with 
radius $R_1(t)<R_2(t)$ which expands exponentially while the 
larger apparent horizon at $R_2(t)$ covering it shrinks.

\item At $t=t_*$ the singularity and the apparent horizon meet at the 
common radius 
\be
R_* = \frac{1}{a} \sqrt{4a^2m^2 +1} >2m \,.
\ee

\item For $t>t_*$ it is $R_2(t)<R_1(t) $ and no apparent horizon exists. 
The singularity is naked. 

\end{itemize}

The behaviour of $R_1(t)$ and $R_2(t)$ is illustrated in 
Fig.~\ref{fig:f(r)=-r^2} for the parameter values $a=1/2, 1$, and $ 3/2$ 
(where all parameters are in units of $m$). A larger $a$ leads to smaller 
$R_*$ and $t_*$.

From Eq.~\eqref{horizon1}, it is clear that the apparent horizon is the 
surface of equation
\be
\psi(t,r) \equiv  r-\frac{\mbox{e}^{-at/2} }{a \sqrt{ 
2\sinh (at)}} =0 \,;
\ee
the normal vector to this surface is 
\begin{equation}
N_{\mu} \equiv 
\nabla_\mu \psi = \frac{\mbox{e}^{at/2}}{2 \sqrt{2} \sinh^{3/2} (at)} \,
\delta^{t}_{\,\,\, \mu}+\delta^{r}_{\,\,\, \mu} 
\end{equation}
and has norm squared 
\begin{equation}
N^\mu N_\mu=\frac{\left(2 \, \mbox{e}^{2 a t}-1\right) \left(2 a m \sqrt{ 
\mbox{e}^{2 a t}-1} -1 \right)}{\left( \mbox{e}^{2 a t}-1\right)^3} \,.
\end{equation}
For $t<t_*$, {\em i.e.}, when this apparent horizon exists, it is 
always $N_c N^c<0$ and this apparent horizon is always spacelike.

\subsection{Physical nature of the apparent horizon for $f(r)=0$ and $f(r)=-r^2$}
\label{sec:4}
\setcounter{equation}{0}

In order to determine the nature of the apparent horizon, we examine the 
expansions of the congruences of outgoing ($+$) and ingoing ($-$) radial 
null geodesics at the apparent horizon. The line element can be written in 
the compact form
\be
d{s}^2= -\frac{dt^2}{A(r)}+ \frac{ \left( \mbox{e}^{2at}-b 
\right)}{A(r)} \, dr^2 + r^2 \, \mbox{e}^{2at} d\Omega_{(2)}^2 \,,
\ee
where $b=0$ for $ f(r)=0$ and $b=1$ for $f(r)=-r^2$. 

Let $\lambda$ be an affine parameter along 
the radial null geodesics and let an overdot denote differentiation with 
respect to $\lambda$. The four-tangents  $ 
\ell^{\mu}_{(\pm)} = \left( \dot{t}, \dot{r}, 0, 0 \right)$ to these 
radial geodesics satisfy the normalization $\ell^c \ell_c=0$, or 
\begin{equation}
\dot{r}^2\left( \mbox{e}^{2at}-b\right)-\dot{t}^2 =0 \,.
\end{equation} 
By introducing the new time coordinate $\bar{t}$ defined by 
\begin{equation}
	d\bar{t}=\frac{dt}{\sqrt{ \mbox{e}^{2at}-b}} \,,
\end{equation}
the line element~\eqref{choice} along the radial null geodesics is  
written in the  conformally flat form
\begin{eqnarray}
d{s}^2\Big|_{d\vartheta=d\varphi=0} &=& \frac{ \left( \mbox{e}^{2at} 
-b \right) }{A(r)} \left(-d\bar{t}^2+dr^2\right) \\
&&\nonumber\\ 
& = & -\frac{ \left( \mbox{e}^{2at} 
-b \right) }{A(r)} \, dudv 
\end{eqnarray}
where, in the last line, we have introduced the null coordinates 
$u \equiv \bar{t}-r $, $v \equiv \bar{t}+r$.

The ingoing and outgoing radial null geodesics, described by  
$v =$~const. and $u =$~const., respectively, have four-tangents 
\begin{equation}
 \ell^{(+)}_{\alpha} \equiv  
-\partial_{\alpha}u = -\frac{1}{\sqrt{ \mbox{e}^{2at}-b}} \, 
\delta^{0}_{\,\,\,\alpha}+\delta^{1}_{\,\,\,\alpha}
\end{equation}
and
\begin{equation}
\ell^{(-)}_{\alpha} \equiv -\partial_{\alpha}v = -\frac{1}{\sqrt{ 
\mbox{e}^{2at}-b}} \, \delta^{0}_{\,\,\,\alpha} -\delta^{1}_{\,\,\,\alpha} 
\end{equation}
in coordinates $\left( t,r, \vartheta, \varphi \right)$, while 
\be
g^{\alpha \beta} \, \ell^{(+)}_\alpha \, \ell^{(-)}_\beta = 
-\frac{2A(r)}{ \mbox{e}^{2at}-b} 
\,.
\ee
In compact form, we have 
\be
\ell_\alpha ^{(\pm)} =-\frac{ \delta ^0_{\,\,\, \alpha}}{ \sqrt{ 
\mbox{e}^{2at} -b }} \pm  \delta ^1 _{\,\,\, \alpha}  \,.
\ee 
The general expression of the expansion scalars of the congruences of 
outgoing and ingoing  radial null geodesics is 
\be
\Theta_{(\pm)} =  
 \left[ g^{ab} - \frac{\ell^a_{(+)} \, \ell^b_{(-)} 
+ \ell^a_{(-)} \,  \ell^b_{(+)}  }{ g^{ab}  \, \ell^{(+)}_a \,  
\ell^{(-)}_b }\right] \nabla_a \, \ell_b^{(\pm)} \,.
\ee
The third term in the right-hand side vanishes if $ \ell^c_{(\pm)}$ is 
globally null; the second term vanishes if the null geodesics are affinely 
parameterized, which we assume here, leaving
\be
\Theta_{(\pm)} = \nabla_c \, \ell^c_{(\pm)} \,.
\ee
We then have
\begin{eqnarray}
\Theta_{(\pm)} &=& \frac{1}{\sqrt{-g}} \, \partial_{\mu} \left( \sqrt{-g} 
\, \ell^{\mu}_{(\pm)} \right) \nonumber\\
&&\nonumber\\ 
&=& \frac{A}{r^2 \, \mbox{e}^{2at} \sin\vartheta \, 
\sqrt{ \mbox{e}^{2at}-b } } \, \partial_{\mu} \left(
r^2 \, \mbox{e}^{2at} \sin\vartheta \, \delta^{\mu}_ 0 \pm \frac{ 
\delta^{\mu}_1  r^2 \, \mbox{e}^{2at} \sin\vartheta }{ \sqrt{ 
\mbox{e}^{2at}-b} } \right) \nonumber\\ 
&&\nonumber\\ 
&=& \frac{A}{r^2 \, \mbox{e}^{2at} \sqrt{ \mbox{e}^{2at}-b}} \left[
\partial_t \left(r^2 \, \mbox{e}^{2at} \right) \pm \frac{ 
\partial_r(r^2) \, \mbox{e}^{2at}  }{ 
\sqrt{ \mbox{e}^{2at}-b}} \right] \nonumber\\
&&\nonumber\\
&=& \frac{2A(r)}{ r \, \sqrt{ \mbox{e}^{2at}-b }  }  \left( ar \pm 
\frac{1}{\sqrt{ \mbox{e}^{2at} -b} } \right) \,.
\end{eqnarray}
Assuming $a>0$, on the apparent horizon of radius $ r_{H}(t)= \left( 
|a|\sqrt{ \mbox{e}^{2at}-b} \, \right)^{-1} $ the expansions assume the 
values
\begin{eqnarray}
\Theta_{(\pm)}( r_\mathrm{H}) &=& \nabla_c \, \ell^c_{(\pm)} 
\Big|_{r_\mathrm{H}} =  \frac{2a A(r_\mathrm{H}) }{ 
\sqrt{ \mbox{e}^{2at}-b } } \, \left( 1 \pm 1\right) \\
&&\nonumber\\
&=& \frac{2a}{\sqrt{ \mbox{e}^{2at}-b} } \left( 1- 2am \sqrt{ 
\mbox{e}^{2at}-b} \right) \, \left( 1\pm 1 \right) \,.
\end{eqnarray}    
If $r_{\rm H}(t) > 2m$ ({\em i.e.}, $A(r_\mathrm{H} (t)) >0$), 
in both cases  $b=0$ (or $f(r)=0$) and $b=1$ (or $f(r)=-r^2$) it is $\Theta_{(+)}(r_\mathrm{H})>0$ and 
$\Theta_{(-)}(r_\mathrm{H}) =0 $.  Furthermore, the Lie derivative of the expansion of ingoing radial null geodesics with respect to $\ell_{(+)}^c$ yields
\begin{equation}
\pounds_{\ell^{(+)}} \, \Theta_{(-)} =  
\, \ell_{(+)}^\alpha \, \partial_\alpha \Theta_{(-)} = 2a^2 A^2(r_{\rm H}) \,  
\frac{ \left( 2 \, \mbox{e}^{2at}-b \right)}{\left( 
\mbox{e}^{2at}-b\right)^2}>0 \,,
\end{equation}
which means that the apparent horizon is a cosmological horizon. 
Therefore, the Fonarev solution \eqref{Fonarev metr}, which describes a 
wormhole, disforms into a spacetime with a cosmological horizon for $f(r)=0$ and $f(r)=-r^2$.

\section{Conclusions}
\label{sec:5}
\setcounter{equation}{0}

While most of the solutions of Horndeski gravity are stationary, the 
ones obtained here by means of disformal transformations are dynamical. 
The change in the physical nature of spherically symmetric objects (black 
holes, wormholes, white holes, naked singularities) under static conformal 
\cite{Faraoni:2015paa,Hammad:2018ldj} and disformal 
\cite{BenAchour:2020wiw,Achour:2021pla,Faraoni:2021gdl} transformations 
has been studied in previous literature, but no general theorems are 
available in the time-dependent case. Here we have considered 
time-dependent disformal transformations of a specific seed, a special 
case of the Fonarev scalar field spacetime of GR \cite{Fonarev:1994xq}. 
The disformation of a white hole or naked singularity (depending on the 
values of the parameters) does not produce a black hole, which is similar 
to the results proved for static transformations 
\cite{BenAchour:2020wiw,Achour:2021pla,Faraoni:2021gdl}.

The spacetimes thus generated are interesting. For $f(r)=-r^2$ we have a 
genuine disformal transformation that originates a geometry with a null 
singularity (reminiscent of thunderbolt singularities 
\cite{Hawking:1992ti, Ishibashi:2002ac,Misonoh:2015nwa}) which expands and 
meets a contracting spacelike apparent horizon. The latter then disappears 
from the spacetime. The expanding singularity is obtained from the 
disformal transformation of a horizon, and it is well-known that apparent 
horizons tend to appear and disappear in pairs 
\cite{Husain:1994uj,Booth:2005qc,Nielsen:2005af, Faraoni:2015ula}, but we 
are not aware of similar phenomenology in the literature, where a 
dynamical singularity annihilates an apparent horizon. The lesson is that, 
in time-dependent dynamical solutions of Horndeski gravity, one should 
not only expect to find phenomenology of dynamical apparent horizons 
familiar from GR and ``old'' scalar-tensor gravity, but also phenomenology 
not encountered before. We will look further for new behaviours of 
apparent horizons and dynamical singularities in future works.


\begin{acknowledgments} 

A.G.~is supported by the European Union's Horizon 2020 research and 
innovation programme under the Marie Sk\l{}odowska-Curie Actions (grant 
agreement No.~895648). The work of A.G~has also been carried out 
in the framework of the activities of the Italian National Group of 
Mathematical Physics [Gruppo Nazionale per la Fisica Matematica (GNFM), 
Istituto Nazionale di Alta Matematica (INdAM)]. V.F. is supported by the 
Natural Sciences \& Engineering Research Council of Canada (Grant 
2016-03803). R.S~and F.S~would like to thank the Iran National Science 
Foundation (INSF) for supporting this research under grant number 
99000365. F.S.~is grateful to the University of Tehran for supporting 
this work under a grant provided by the university research council.

\end{acknowledgments}


\begin{thebibliography}{99}

\bibitem{Capozziello:2003tk} S.~Capozziello, S.~Carloni and A.~Troisi, 
``Quintessence without scalar fields,'' Recent Res. Dev. Astron. 
Astrophys. \textbf{1}, 625 (2003) [arXiv:astro-ph/0303041 [astro-ph]].

\bibitem{Carroll:2003wy} S.~M.~Carroll, V.~Duvvuri, M.~Trodden and 
M.~S.~Turner, ``Is cosmic speed - up due to new gravitational physics?,'' 
Phys. Rev. D \textbf{70}, 043528 (2004) 
[arXiv:astro-ph/0306438 [astro-ph]].

\bibitem{Brans:1961sx} C.~Brans and R.~H.~Dicke, ``Mach's principle 
and a relativistic theory of gravitation'', Phys. Rev.  
\textbf{124}, 
925-935 (1961) 

\bibitem{Bergmann:1968ve} P.~G.~Bergmann, ``Comments on the scalar 
tensor theory'', Int. J. Theor. Phys. \textbf{1}, 25-36 (1968) 

\bibitem{Nordtvedt:1968qs} K.~Nordtvedt, ``Equivalence Principle for 
Massive Bodies. 2. Theory'', Phys. Rev. \textbf{169}, 1017-1025 
(1968) doi:10.1103/PhysRev.169.1017.

\bibitem{Wagoner:1970vr} R.~V.~Wagoner, ``Scalar tensor theory and 
gravitational waves'', Phys. Rev. D \textbf{1}, 3209-3216 (1970) 
doi:10.1103/PhysRevD.1.3209.

\bibitem{Nordtvedt:1970uv} K.~Nordtvedt, Jr., ``PostNewtonian metric 
for a general class of scalar tensor gravitational theories and 
observational consequences'', Astrophys. J. \textbf{161}, 1059-1067 
(1970) doi:10.1086/150607.

\bibitem{AmendolaTsujikawabook} L. Amendola and S. 
Tsujikawa, {\em Dark Energy, Theory and Observations} 
(Cambridge University Press, Cambridge, England, 2010).

\bibitem{Faraoni:2004pi} V.~Faraoni, {\em Cosmology in Scalar-Tensor 
Gravity} (Kluwer Academic, Dordrecht, 2004) doi:10.1007/978-1-4020-1989-0

\bibitem{Faraoni:2010pgm} V.~Faraoni and S.~Capozziello, {\em Beyond 
Einstein Gravity: A Survey of Gravitational Theories for Cosmology and 
Astrophysics}, (Springer, New York, 2010) doi:10.1007/978-94-007-0165-6

\bibitem{Sotiriou:2008rp} T.~P.~Sotiriou and V.~Faraoni, ``$f(R)$ Theories 
of Gravity,'' Rev. Mod. Phys. \textbf{82} (2010) 451  
doi:10.1103/RevModPhys.82.451 
[arXiv:0805.1726 [gr-qc]].

\bibitem{DeFelice:2010aj} A.~De Felice and S.~Tsujikawa, ``$f(R)$  
theories,'' Living Rev. Rel. \textbf{13} (2010) 3 
[arXiv:1002.4928 [gr-qc]].

\bibitem{Nojiri:2010wj}
S.~Nojiri and S.~D.~Odintsov,
``Unified cosmic history in modified gravity: from F(R) theory to Lorentz 
non-invariant models,''
Phys. Rept. \textbf{505}, 59-144 (2011)
doi:10.1016/j.physrep.2011.04.001
[arXiv:1011.0544 [gr-qc]].

\bibitem{Deffayet:2009wt}
C.~Deffayet, G.~Esposito-Farese and A.~Vikman,
``Covariant Galileon,''
Phys. Rev. D \textbf{79}, 084003 (2009)
doi:10.1103/PhysRevD.79.084003
[arXiv:0901.1314 [hep-th]].

\bibitem{Deffayet:2009mn}
C.~Deffayet, S.~Deser and G.~Esposito-Farese,
``Generalized Galileons: All scalar models whose curved background 
extensions maintain second-order field equations and stress-tensors,''
Phys. Rev. D \textbf{80}, 064015 (2009)
doi:10.1103/PhysRevD.80.064015
[arXiv:0906.1967 [gr-qc]].

\bibitem{Deffayet:2011gz}
C.~Deffayet, X.~Gao, D.~A.~Steer and G.~Zahariade,
``From k-essence to generalised Galileons,''
Phys. Rev. D \textbf{84}, 064039 (2011)
doi:10.1103/PhysRevD.84.064039
[arXiv:1103.3260 [hep-th]].

\bibitem{Horndeski:1974wa}
G.~W.~Horndeski,
``Second-order scalar-tensor field equations in a four-dimensional 
space,''
Int. J. Theor. Phys. \textbf{10}, 363-384 (1974)
doi:10.1007/BF01807638

\bibitem{Gleyzes:2014dya}
J.~Gleyzes, D.~Langlois, F.~Piazza and F.~Vernizzi,
``Healthy theories beyond Horndeski,''
Phys. Rev. Lett. \textbf{114}, no.21, 211101 (2015)
doi:10.1103/PhysRevLett.114.211101
[arXiv:1404.6495 [hep-th]].

\bibitem{Gleyzes:2014qga}
J.~Gleyzes, D.~Langlois, F.~Piazza and F.~Vernizzi,
``Exploring gravitational theories beyond Horndeski,''
JCAP \textbf{02}, 018 (2015)
doi:10.1088/1475-7516/2015/02/018
[arXiv:1408.1952 [astro-ph.CO]].

\bibitem{Langlois:2015cwa}
D.~Langlois and K.~Noui,
``Degenerate higher derivative theories beyond Horndeski: evading the 
Ostrogradski instability,''
JCAP \textbf{02}, 034 (2016)
doi:10.1088/1475-7516/2016/02/034
[arXiv:1510.06930 [gr-qc]].

\bibitem{Langlois:2015skt}
D.~Langlois and K.~Noui,
``Hamiltonian analysis of higher derivative scalar-tensor theories,''
JCAP \textbf{07}, 016 (2016)
doi:10.1088/1475-7516/2016/07/016
[arXiv:1512.06820 [gr-qc]].

\bibitem{BenAchour:2016cay}
J.~Ben Achour, D.~Langlois and K.~Noui,
``Degenerate higher order scalar-tensor theories beyond Horndeski and 
disformal transformations,''
Phys. Rev. D \textbf{93}, no.12, 124005 (2016)
doi:10.1103/PhysRevD.93.124005
[arXiv:1602.08398 [gr-qc]].

\bibitem{Crisostomi:2016czh}
M.~Crisostomi, K.~Koyama and G.~Tasinato,
``Extended Scalar-Tensor Theories of Gravity,''
JCAP \textbf{04}, 044 (2016)
doi:10.1088/1475-7516/2016/04/044
[arXiv:1602.03119 [hep-th]].

\bibitem{Motohashi:2016ftl}
H.~Motohashi, K.~Noui, T.~Suyama, M.~Yamaguchi and D.~Langlois,
``Healthy degenerate theories with higher derivatives,''
JCAP \textbf{07}, 033 (2016)
doi:10.1088/1475-7516/2016/07/033
[arXiv:1603.09355 [hep-th]].

\bibitem{BenAchour:2016fzp}
J.~Ben Achour, M.~Crisostomi, K.~Koyama, D.~Langlois, K.~Noui and 
G.~Tasinato,
``Degenerate higher order scalar-tensor theories beyond Horndeski up to 
cubic order,''
JHEP \textbf{12}, 100 (2016)
doi:10.1007/JHEP12(2016)100
[arXiv:1608.08135 [hep-th]].

\bibitem{Crisostomi:2017aim}
M.~Crisostomi, R.~Klein and D.~Roest,
``Higher Derivative Field Theories: Degeneracy Conditions and Classes,''
JHEP \textbf{06}, 124 (2017)
doi:10.1007/JHEP06(2017)124
[arXiv:1703.01623 [hep-th]].

\bibitem{Langlois:2018dxi}
D.~Langlois,
``Dark energy and modified gravity in degenerate higher-order 
scalar\textendash{}tensor (DHOST) theories: A review,''
Int. J. Mod. Phys. D \textbf{28}, no.05, 1942006 (2019)
doi:10.1142/S0218271819420069
[arXiv:1811.06271 [gr-qc]].

\bibitem{Langlois:2017mdk}
D.~Langlois,
``Degenerate Higher-Order Scalar-Tensor (DHOST) theories,''
[arXiv:1707.03625 [gr-qc]].

\bibitem{Faraoni:2021nhi} V.~Faraoni, A.~Giusti and B.~H.~Fahim, 
``Spherical inhomogeneous solutions of Einstein and 
scalar\textendash{}tensor gravity: A map of the land,'' Phys. Rept. 
\textbf{925}, 1-58 (2021) doi:10.1016/j.physrep.2021.04.003 
[arXiv:2101.00266 [gr-qc]].

\bibitem{BenAchour:2020wiw}
J.~Ben Achour, H.~Liu and S.~Mukohyama,
``Hairy black holes in DHOST theories: Exploring disformal transformation 
as a solution-generating method,''
JCAP \textbf{02}, 023 (2020)
doi:10.1088/1475-7516/2020/02/023
[arXiv:1910.11017 [gr-qc]].

\bibitem{Faraoni:2021gdl}
V.~Faraoni and A.~Leblanc,
``Disformal mappings of spherical DHOST geometries,''
JCAP \textbf{08}, 037 (2021)
doi:10.1088/1475-7516/2021/08/037
[arXiv:2107.03456 [gr-qc]].

\bibitem{Faraoni:2015paa}
V.~Faraoni, A.~Prain and A.~F.~Zambrano Moreno, ``Black holes and 
wormholes subject 
to conformal mappings,'' Phys. Rev. D \textbf{93}, no.2, 024005 (2016)
doi:10.1103/PhysRevD.93.024005
[arXiv:1509.04129 [gr-qc]].

\bibitem{Hammad:2018ldj}
F.~Hammad,
``Revisiting black holes and wormholes under Weyl transformations,''
Phys. Rev. D \textbf{97}, no.12, 124015 (2018)
doi:10.1103/PhysRevD.97.124015
[arXiv:1806.01388 [gr-qc]].

\bibitem{Achour:2021pla}
J.B.~Achour, A.~De Felice, M.A.~Gorji, S.~Mukohyama and 
M.C.~Pookkillath,
``Disformal map and Petrov classification in modified gravity,''
[arXiv:2107.02386 [gr-qc]].

\bibitem{Babichev:2012re}
E.~Babichev and G.~Esposito-Far\`ese,
``Time-Dependent Spherically Symmetric Covariant Galileons,''
Phys. Rev. D \textbf{87}, 044032 (2013)
doi:10.1103/PhysRevD.87.044032
[arXiv:1212.1394 [gr-qc]].

\bibitem{Anabalon:2013oea}
A.~Anabalon, A.~Cisterna and J.~Oliva,
``Asymptotically locally AdS and flat black holes in Horndeski theory,''
Phys. Rev. D \textbf{89}, 084050 (2014)
doi:10.1103/PhysRevD.89.084050
[arXiv:1312.3597 [gr-qc]].

\bibitem{Babichev:2013cya}
E.~Babichev and C.~Charmousis,
``Dressing a black hole with a time-dependent Galileon,''
JHEP \textbf{08}, 106 (2014)
doi:10.1007/JHEP08(2014)106
[arXiv:1312.3204 [gr-qc]].

\bibitem{Charmousis:2014zaa}
C.~Charmousis, T.~Kolyvaris, E.~Papantonopoulos and M.~Tsoukalas,
``Black Holes in Bi-scalar Extensions of Horndeski Theories,''
JHEP \textbf{07}, 085 (2014)
doi:10.1007/JHEP07(2014)085
[arXiv:1404.1024 [gr-qc]].

\bibitem{Kobayashi:2014eva}
T.~Kobayashi and N.~Tanahashi,
``Exact black hole solutions in shift symmetric scalar\textendash{}tensor 
theories,''
PTEP \textbf{2014}, 073E02 (2014)
doi:10.1093/ptep/ptu096
[arXiv:1403.4364 [gr-qc]].

\bibitem{Babichev:2016kdt}
E.~Babichev and G.~Esposito-Farese,
``Cosmological self-tuning and local solutions in generalized Horndeski 
theories,''
Phys. Rev. D \textbf{95}, no.2, 024020 (2017)
doi:10.1103/PhysRevD.95.024020
[arXiv:1609.09798 [gr-qc]].

\bibitem{Motohashi:2018wdq}
H.~Motohashi and M.~Minamitsuji,
``General Relativity solutions in modified gravity,''
Phys. Lett. B \textbf{781}, 728-734 (2018)
doi:10.1016/j.physletb.2018.04.041
[arXiv:1804.01731 [gr-qc]].

\bibitem{Babichev:2017guv}
E.~Babichev, C.~Charmousis and A.~Leh\'ebel,
``Asymptotically flat black holes in Horndeski theory and beyond,''
JCAP \textbf{04}, 027 (2017)
doi:10.1088/1475-7516/2017/04/027
[arXiv:1702.01938 [gr-qc]].

\bibitem{Babichev:2013cya}
E.~Babichev and C.~Charmousis,
``Dressing a black hole with a time-dependent Galileon,''
JHEP \textbf{08}, 106 (2014)
doi:10.1007/JHEP08(2014)106
[arXiv:1312.3204 [gr-qc]].

\bibitem{Anson:2020trg}
T.~Anson, E.~Babichev, C.~Charmousis and M.~Hassaine,
``Disforming the Kerr metric,''
JHEP \textbf{01}, 018 (2021)
doi:10.1007/JHEP01(2021)018
[arXiv:2006.06461 [gr-qc]].

\bibitem{BenAchour:2020fgy}
J.~Ben Achour, H.~Liu, H.~Motohashi, S.~Mukohyama and K.~Noui,
``On rotating black holes in DHOST theories,''
JCAP \textbf{11}, 001 (2020)
doi:10.1088/1475-7516/2020/11/001
[arXiv:2006.07245 [gr-qc]].

\bibitem{Chatzifotis:2021hpg}
N.~Chatzifotis, E.~Papantonopoulos and C.~Vlachos,
``Disformal transition of a black hole to a wormhole in scalar-tensor 
Horndeski theory,''
Phys. Rev. D \textbf{105}, no.6, 064025 (2022)
doi:10.1103/PhysRevD.105.064025
[arXiv:2111.08773 [gr-qc]].

\bibitem{Waldbook} R.~M. Wald, {\em General Relativity} (Chicago 
University Press, Chicago, 1984).

\bibitem{Bekenstein:1992pj}
J.~D.~Bekenstein,
``The Relation between physical and gravitational geometry,''
Phys. Rev. D \textbf{48}, 3641-3647 (1993)
doi:10.1103/PhysRevD.48.3641
[arXiv:gr-qc/9211017 [gr-qc]].

\bibitem{Ezquiaga:2017ner}
J.~M.~Ezquiaga, J.~Garc\'\i{}a-Bellido and M.~Zumalac\'arregui,
``Field redefinitions in theories beyond Einstein gravity using the 
language of differential forms,''
Phys. Rev. D \textbf{95}, no.8, 084039 (2017)
doi:10.1103/PhysRevD.95.084039
[arXiv:1701.05476 [hep-th]].

\bibitem{Zumalacarregui:2010wj}
M.~Zumalacarregui, T.~S.~Koivisto, D.~F.~Mota and P.~Ruiz-Lapuente,
``Disformal Scalar Fields and the Dark Sector of the Universe,''
JCAP \textbf{05}, 038 (2010)
doi:10.1088/1475-7516/2010/05/038
[arXiv:1004.2684 [astro-ph.CO]].

\bibitem{Zumalacarregui:2013pma}
M.~Zumalac\'arregui and J.~Garc\'\i{}a-Bellido,
``Transforming gravity: from derivative couplings to matter to 
second-order 
scalar-tensor theories beyond the Horndeski Lagrangian,''
Phys. Rev. D \textbf{89}, 064046 (2014)
doi:10.1103/PhysRevD.89.064046
[arXiv:1308.4685 [gr-qc]].

\bibitem{Wald:1991zz}
R.~M.~Wald and V.~Iyer,
``Trapped surfaces in the Schwarzschild geometry and cosmic censorship,''
Phys. Rev. D \textbf{44}, R3719-R3722 (1991)
doi:10.1103/PhysRevD.44.R3719

\bibitem{Schnetter:2005ea}
E.~Schnetter and B.~Krishnan,
``Non-symmetric trapped surfaces in the Schwarzschild and Vaidya 
spacetimes,''
Phys. Rev. D \textbf{73}, 021502 (2006)
doi:10.1103/PhysRevD.73.021502
[arXiv:gr-qc/0511017 [gr-qc]].

\bibitem{Faraoni:2016xgy}
V.~Faraoni, G.~F.~R.~Ellis, J.~T.~Firouzjaee, A.~Helou and I.~Musco,
``Foliation dependence of black hole apparent horizons in spherical 
symmetry,''
Phys. Rev. D \textbf{95}, no.2, 024008 (2017)
doi:10.1103/PhysRevD.95.024008
[arXiv:1610.05822 [gr-qc]].

\bibitem{Husain:1994uj} V.~Husain, E.~A.~Martinez and D.~Nu\~nez, ``Exact 
solution for scalar field collapse,'' Phys. Rev. D \textbf{50}, 3783-3786 
(1994) doi:10.1103/PhysRevD.50.3783 [arXiv:gr-qc/9402021 [gr-qc]].

\bibitem{Booth:2005qc}
I.~Booth, ``Black hole boundaries,''
Can. J. Phys. \textbf{83}, 1073-1099 (2005)
doi:10.1139/p05-063
[arXiv:gr-qc/0508107 [gr-qc]].

\bibitem{Nielsen:2005af} A.~B.~Nielsen and M.~Visser, ``Production and 
decay of evolving horizons,'' Class. Quant. Grav. \textbf{23}, 4637-4658 
(2006) doi:10.1088/0264-9381/23/14/006 [arXiv:gr-qc/0510083 [gr-qc]].

\bibitem{Abreu:2010ru} G.~Abreu and M.~Visser, ``Kodama time: 
Geometrically preferred foliations of spherically symmetric spacetimes,'' 
Phys. Rev. D \textbf{82}, 044027 (2010) doi:10.1103/PhysRevD.82.044027 
[arXiv:1004.1456 [gr-qc]].

\bibitem{Faraoni:2015ula} V.~Faraoni, {\em Cosmological and Black Hole 
Apparent Horizons}, Lect. Notes Phys. \textbf{907} (Springer, New York, 
2015) doi:10.1007/978-3-319-19240-6

\bibitem{Fonarev:1994xq} O.~A.~Fonarev, ``Exact Einstein scalar field 
solutions for formation of black holes in a cosmological setting,'' Class. 
Quant. Grav. \textbf{12}, 1739-1752 (1995) doi:10.1088/0264-9381/12/7/016 
[arXiv:gr-qc/9409020 [gr-qc]].

\bibitem{Maeda:2007bu} H.~Maeda, ``Global structure and physical 
interpretation of the Fonarev solution for a scalar field with exponential 
potential,'' [arXiv:0704.2731 [gr-qc]].

\bibitem{Fisher:1948yn} I.~Z.~Fisher, ``Scalar mesostatic field with 
regard for gravitational effects,'' Zh. Eksp. Teor. Fiz. \textbf{18}, 
636-640 (1948) [arXiv:gr-qc/9911008 [gr-qc]].

\bibitem{Janis:1968zz}
A.~I.~Janis, E.~T.~Newman and J.~Winicour,
``Reality of the Schwarzschild Singularity,''
Phys. Rev. Lett. \textbf{20}, 878-880 (1968)
doi:10.1103/PhysRevLett.20.878

\bibitem{Wyman:1981bd}
M.~Wyman,
``Static Spherically Symmetric Scalar Fields in General Relativity,''
Phys. Rev. D \textbf{24}, 839-841 (1981)
doi:10.1103/PhysRevD.24.839

\bibitem{Faraoni:2017afs} V.~Faraoni and S.~D.~Belknap-Keet, ``New 
inhomogeneous universes in scalar-tensor and $f(R)$ gravity,'' Phys. Rev. 
D \textbf{96}, no.4, 044040 (2017) doi:10.1103/PhysRevD.96.044040 
[arXiv:1705.05749 [gr-qc]].

\bibitem{Schleich:2009uj}
K.~Schleich and D.~M.~Witt,
``A simple proof of Birkhoff's theorem for cosmological constant,''
J. Math. Phys. \textbf{51}, 112502 (2010)
doi:10.1063/1.3503447
[arXiv:0908.4110 [gr-qc]].

\bibitem{Hawking:1992ti}
S.~W.~Hawking and J.~M.~Stewart,
``Naked and thunderbolt singularities in black hole evaporation,''
Nucl. Phys. B \textbf{400}, 393-415 (1993)
doi:10.1016/0550-3213(93)90410-Q
[arXiv:hep-th/9207105 [hep-th]].

\bibitem{Ishibashi:2002ac}
A.~Ishibashi and A.~Hosoya,
``Naked singularity and thunderbolt,''
Phys. Rev. D \textbf{66}, 104016 (2002)
doi:10.1103/PhysRevD.66.104016
[arXiv:gr-qc/0207054 [gr-qc]].

\bibitem{Misonoh:2015nwa}
Y.~Misonoh and K.~i.~Maeda,
``Black Holes and Thunderbolt Singularities with Lifshitz Scaling 
Terms,''
Phys. Rev. D \textbf{92}, no.8, 084049 (2015)
doi:10.1103/PhysRevD.92.084049
[arXiv:1509.01378 [gr-qc]].

\end{thebibliography}
\end{document}